\documentclass[11pt,nofootinbib]{revtex4}
\usepackage{amsfonts,amssymb,amsmath,graphicx}
\usepackage[latin1]{inputenc}
\usepackage[T1]{fontenc}
\usepackage{subfigure}
\usepackage{fancyhdr}
\usepackage{mathrsfs}
\usepackage{color,psfrag}

\def\dac{\displaystyle\frac}

\def\({\left(}
\def\){\right)}

\begin{document}

\title{\textbf{From Big to Little Rip in modified F(R,G) gravity}}

\author{Makarenko Andrey N.\footnote{Electronic address:
andre@tspu.edu.ru}, Obukhov Valery V.\footnote{Electronic address:
rector@tspu.edu.ru}, Kirnos Ilya V.\footnote{Electronic address:
kiv@keva.tusur.ru}} \affiliation{Department of Theoretical Physics,
Tomsk State Pedagogical University, Tomsk, 634041 Russia}

\begin{abstract}

We discuss the cosmological reconstruction in modified Gauss-Bonnet (GB) gravity.
It is demonstrated that the modified GB gravity may describe the most interesting features
of late-time cosmology.
We derive explicit
form of effective phantom cosmological models
ending by the finite-time future singularity (Big Rip) and without singularities in the future (Little Rip).

\end{abstract}

\maketitle

\section{INTRODUCTION}

Recent observational data indicate that our Universe is currently in accelerated phase \cite{Dat}. Since the discovery of the acceleration, a large number of possible
mechanisms have been proposed to explain the origin of the dark energy.
General
Relativity in its standard form can not explain the accelerated expansion without extra terms or components, which
have been gathered under the name of dark energy.
There are various proposals to construct an acceptable dark energy model, such as: a scalar field with quintessence-like or phantom-like behavior \cite{Eli1}, spinor, (non-)abelian vector
theory, cosmological constant, fluid with a complicated equation of state, and higher dimensions \cite{mak1}.
As a result, we have a number of competing scenarios
 which to describe late-time acceleration.  Working in framework of one of the above proposals, we are forced to introduce
the following extra cosmological components: inflaton, a dark component and dark matter.

We concentrate on the
modified gravity \cite{N1,N11}, which represents a classical generalization of general
relativity (modifications
of the Hilbert-Einstein action by introducing different functions of the Ricci scalar \cite{N11,fr} or Gauss-Bonnet invariant \cite{gb,gb1}), should consistently describe the early-time inflation and late-time acceleration, without the introduction of
any other dark component.

In the present work we will focus on the Gauss-Bonnet gravity, where the gravitational action includes functions of the Gauss-Bonnet invariant
\begin{equation}
 G=R^2-4R_{\mu\nu}R^{\mu\nu}+R_{\mu\nu\lambda\sigma}R^{\mu\nu\lambda\sigma}\ .
\label{I2}
\end{equation}
It is interesting, that the theory of this type is closely related to higher order string curvature corrections \cite{N65}.
The starting action is
\begin{equation}
\label{GBany1}
S=\int d^4 x \sqrt{-g}\left[ \frac{R}{2\kappa^2} - \frac{1}{2}\partial_\mu \phi
\partial^\mu \phi - V(\phi) - \xi_1(\phi) G \right]\ .
\end{equation}

Here $G$ is the Gauss-Bonnet invariant (\ref{I2})
and the scalar field $\phi$ is canonical in (\ref{GBany1}).

One may redefine the scalar field $\phi$ by
$\phi=\epsilon
\varphi$.
The action takes the following form
\begin{equation}
\label{GBany41}
S=\int d^4 x \sqrt{-g}\left[ \frac{R}{2\kappa^2} -
\frac{\epsilon^2}{2}\partial_\mu \phi \partial^\mu \phi
    - \tilde V(\varphi) - \tilde\xi_1(\varphi) G \right]\ .
\end{equation}
Here
\begin{equation}
\label{GBany42}
\tilde V(\varphi) \equiv V(\epsilon\varphi)\ ,\quad
\tilde\xi_1(\varphi) \equiv \xi_1(\epsilon\varphi)\ .
\end{equation}
If a proper limit $\epsilon\to 0$ exists, the action (\ref{GBany41}) reduces
to
\begin{equation}
\label{GBany43}
S=\int d^4 x \sqrt{-g}\left[ \frac{R}{2\kappa^2} - \tilde V(\varphi) -
\tilde\xi_1(\varphi) G \right]\ .
\end{equation}
Then  $\varphi$ is an auxiliary field. By variation of
$\varphi$, we find
\begin{equation}
\label{GBany44}
0={\tilde V}'(\varphi) - {\tilde\xi_1}'(\varphi)G\ ,
\end{equation}
which may be solved with respect to $\varphi$ as
\begin{equation}
\label{GBany45}
\varphi=\Phi(G)\ .
\end{equation}
Substituting (\ref{GBany46}) into the action (\ref{GBany43}), we get for the
$F(G)$-gravity
\begin{equation}
\label{GBany46}
S=\int d^4 x \sqrt{-g}\left[ \frac{R}{2\kappa^2} - F(G)\right]\ ,\quad
F(G)\equiv \tilde V\left(\Phi(G)\right) - \tilde\xi_1\left(\Phi(G)\right) G\ .
\end{equation}

On the other hand, the action \ref{1.1} is associated with non-local models \cite{Eli5}
Let's start with the next action
\begin{equation}
\label{nlGB1}
S=\int d^4 x \sqrt{-g} \left(\frac{R}{2\kappa^2}
 - \frac{\kappa^2}{2\alpha} G\Box^{-1}  G + {\cal L}_{m} \right)\ .
\end{equation}
Here ${\cal L}_m$ is the matter Lagrangian.
By introducing the scalar field $\phi$, one may rewrite the action
(\ref{nlGB1}) in a local form:
\begin{equation}
\label{nlGB3}
S=\int d^4 x \left(\frac{R}{2\kappa^2} - \frac{\alpha}{2\kappa^2}\partial_\mu\phi \partial^\mu \phi
+ \phi  G + {\cal L}_{m} \right)\ .
\end{equation}
In fact, from the $\phi$-equation we find $\phi = - \frac{\kappa^2}{\alpha} \Box^{-1}  G$.
By substituting this expression into (\ref{nlGB3}), we reobtain (\ref{nlGB1}).

Thus, by studying $F(G)$ -  gravity, we are also exploring several other models. Note that in some cases, $F(G)$ or even $F(R,G$) gravity maybe transformed to
$F(R)$ gravity \cite{All}.

The five-year Wilkinson Microwave Anisotropy Probe (WMAP)
data \cite{Dat} give the bounds to the value of the equation of state (EoS) parameter $w_{DE}$, which is the ratio of the pressure
of the dark energy to the energy density of it, in the range of -1.11 < $w_{DE}$ < -0.86. This could be consistent if the
dark energy is a cosmological constant with $w_{DE}$ = -1 and therefore our universe seems to approach asymptotically
de Sitter universe.

Although the accelerating expansions seem to be de Sitter type, the possibility that the current acceleration
could be quintessence type, in which $w_{DE}$ > -1, or phantom type, in which $w_{DE}$ < -1, is not completely excluded.

It is often assumed that the early universe started from the singular point often called
Big Bang. However, if the current (or future) universe enters the quintessence/phantom stage, it may evolve to the
finite-time future singularity depending on the specific model under consideration and the value of the effective EoS
parameter (Big Rip) \cite{Sig}.
An elegant solution to this problem has been recently proposed \cite{frampton11} under the form of the so-called Little Rip cosmology which appears
to be a realistic alternative to the $\Lambda$CDM model.

\section{$[R+f(G)]$ gravity}

Consider the following action (see Ref. ~\cite{gb1}):
\begin{equation}
S=\int d^{4} x \sqrt{-g} \left(\frac{1}{2\kappa^{2}}R+f(G)+L_{m}\right)
\label{GB11}
\end{equation}
where $\kappa^2=8 \pi G_N, \, G_N$ being the Newton constant, and the
Gauss-Bonnet invariant is defined by(\ref{I2}).
By varying the action (\ref{GB11}) over $g_{\mu\nu}$, are obtained:
\begin{eqnarray}
0&=& \frac{1}{2k^2} \left(-R^{\mu\nu}+\frac{1}{2}g^{\mu\nu}R\right)+T^{\mu\nu}+\frac{1}{2} g^{\mu\nu} f(G) - 2 f_{G} R R^{\mu\nu} + 4 f_{G} R^{\mu}_{\alpha} R^{\nu\alpha}
-2f_{G}R^{\mu\alpha\beta\tau}R^{\nu}_{\alpha\beta\tau}-\nonumber\\
&-&4 f_{G} R^{\mu\alpha\beta\nu} R_{\alpha\beta}+2\left(\nabla^\mu \nabla^\nu f_{G} \right) R
-2g^{\mu\nu}(\nabla^{2}f_{G})R-4(\nabla_{\rho}\nabla^{\mu}f_{G})R^{\nu\rho}
\nonumber\\
&-&4(\nabla_{\rho}\nabla^{\nu}f_{G})R^{\mu\rho}+4(\nabla^{2}f_{G})R^{\mu\nu}+4g^{\mu\nu}(\nabla_{\rho}\nabla_{\sigma}f_{G})R^{\rho\sigma}
-4(\nabla_{\alpha}\nabla_{\beta}f_{G})R^{\mu\alpha\nu\beta},\nonumber
\end{eqnarray}
where $f_G=f'(G)$ and $f_{GG}=f''(G)$ ($'=\partial / \partial G$).

We consider a spatially-flat universe
\begin{equation}
ds^{2}=-dt^{2}+a(t)^{2}\sum^{3}_{i=1}(dx^{i})^{2},
\label{FRW1}.
\end{equation}
Here $a(t)$ is the scale factor at cosmological time $t$.
It is easy to obtain the FRW equations
\begin{equation}
0=-\frac{3}{\kappa^{2}}H^{2}+G f_{G}-f(G)-24\dot{G}H^{3}f_{GG}+\rho_{m}, \nonumber
\end{equation}
\begin {equation}
0=8 H^2 \ddot{f}_{G}+16 H (\dot{H}+H^2) \dot{f}_G+\frac{1}{\kappa^2}(2\dot{H}+3H^2)+f-Gf_G+p_m.
\label{GB13}
\end{equation}

Here  $H$ is the Hubble rate ($H=\dot{a}/a$, $\dot{}=\partial / \partial t$),
$\rho_{m}$ is matter energy density:
\begin {equation}
\dot{\rho_{m}}+3H(1+w)\rho_{m}=0\ , \label{GB15}
\end{equation}
while  $G$, its derivative and
$R$ can be defined as functions of the Hubble parameter as
\begin{eqnarray}
G=24(\dot{H}H^{2}+H^{4}), \,\, R=6(\dot{H}+2H^{2}),\nonumber\\
\dot{G}=24 H (\ddot{H}H+2\dot{H}^2+4 H^{2}).\nonumber
\end{eqnarray}

It is easy to see that the second equation in (\ref{GB13})  is a differential consequence on the solution.

The first equation (\ref{GB13}) can thus be expressed as follows
\begin {equation}
\label{FFF1}
0=-\frac{3}{\kappa^{2}}H^{2}+24(\dot{H}H^{2}+H^{4})
f_{G}-f(G)-24^2 H^4 (\ddot{H}H+2\dot{H}^2+4 H^{2})f_{GG}+\rho_{m}\,,
\end{equation}

By selecting certain of the Hubble parameter  $H$ we obtain a differential equation for the function $f$.
This is the method of reconstruction (this method has been implemented in Ref.~\cite{N4} for $f(R)$
gravity and Ref.~\cite{gb1} for $f(G)$ gravity).

\section{$f(R,\,G)$ gravity}

Let us now consider a more general model for a kind of modified Gauss-Bonnet gravity.
This can be described by the following action
\begin{equation}
S=\int d^{4}x\sqrt{-g}\left[\frac{1}{2k^2}F(R,G)+L_{m}\right].
\label{F(R,G)action}
\end{equation}
Varying over $g_{\mu\nu}$ the gravity field equations are obtained \cite{gb1},
\begin{eqnarray}
0=T^{\mu\nu}+\frac{1}{2}g^{\mu\nu}F(G)-2F_{R,G}RR^{\mu\nu}+4F_{G}R^{\mu}_{\rho}R^{\nu\rho}\nonumber\\
-2F_{G}R^{\mu\rho\sigma\tau}R^{\nu}_{\rho\sigma\tau}-4F_{G}R^{\mu\rho\sigma\nu}R_{\rho\sigma}+2(\nabla^{\mu}\nabla^{\nu}F_{G})R
-2g^{\mu\nu}(\nabla^{2}F_{G})R\nonumber\\
-4(\nabla_{\rho}\nabla^{\mu}F_{G})R^{\nu\rho}-4(\nabla_{\rho}\nabla^{\nu}F_{G})R^{\mu\rho}+4(\nabla^{2}F_{G})R^{\mu\nu}+4g^{\mu\nu}(\nabla_{\rho}\nabla_{\sigma}F_{G})R^{\rho\sigma}\nonumber\\
-4(\nabla_{\rho}\nabla_{\sigma}F_{G})R^{\mu\rho\nu\sigma}-F_{R}R^{\mu\nu}+\nabla^{\mu}\nabla^{\nu}F_{R}-g^{\mu\nu}\nabla^{2}F_{R}.
\label{FieldEq}
\end{eqnarray}
In the case of a flat FRW Universe, described by the metric (\ref{FRW}), the first FRW equation yields
\begin {equation}
0=\frac{1}{2}(GF_{G}-F-24H^{3}F_{Gt})+3(\dot{H}+H^{2})F_{R}-3HF_{Rt}+k^2\rho_{m}.
\label{FriedmEq}
\end{equation}

 For simplicity, we consider the following subfamily of functions
\begin{equation}
F(R,G)=f_1(G)+f_2(R)\ .
\label{3.1a}
\end{equation}
Correspondingly, the Friedmann equation (\ref{FriedmEq}) can be split
into two equations, as
\begin{eqnarray}
 0=-24H^3\dot{G}f_{1GG}+ Gf_{1G}-f_1\ , \nonumber\\
0=-3H\dot{R}f_{2RR}+3(\dot{H}+H^2)f_{2R}-\frac{1}{2}f_2 +\kappa^2\rho_m
\label{3.1}
\end{eqnarray}

\section{Little Rip model}

In  Ref.~\cite{LIT1} is proposed Little Rip model on the basis of a viscous fluid.
Using the obtained in this study form the scale factor, we construct a model of the GB gravity.

Consider the scale factor as \cite{LIT1}
\begin{equation} \label{c1}
a(t)=\exp{\alpha \left(e^{\beta t}-1\right)}
\end{equation}

In this case the Hubble parameter has the form
\begin{equation} \label{h1}
H(t)=\alpha\beta e^{\beta t}
\end{equation}

Then the derivative of the Hubble parameter is proportional to itself, and
\begin{equation} \label{gg1}
G=24 H^3\left( \frac{H}{\beta}-\alpha\right).
\end{equation}
FRW equation for the $R+f(G)$ theory gravity takes the form

\begin{equation} \label{FRW11}
0=-\frac{3}{\kappa^2} H^2-f+\frac{9\beta^2 H+19\beta H^2+4H^3}{(3\beta+4H)^2}\frac{df}{d H}-\frac{\beta H^2}{3\beta+4H}\frac{d^2 f}{d H^2}+\rho_m.
\end{equation}

Choose the energy density in the form  $\rho_m=const$ we
obtain the following solution
\begin{eqnarray} \label{sol1}
f(H)&=&-\frac{18 \beta^3 H+3 \beta H^3+3 H^4+2 \beta^2 \left(6 H^2-\kappa^2 \rho_m\right)}{2\beta^2 \kappa^2}+\frac{H^3 (\beta+H) C_1}{1+\beta}+\\
&+&\frac{H C_2 \left(-\beta e^{\frac{H}{\beta}} \left(3 \beta^2+2 \beta H+H^2\right)+H^2 (\beta+H) \text{ExpIntegralEi}\left[\frac{H}{\beta}\right]\right)}{2\beta^3 (1+\beta)} \nonumber
\end{eqnarray}

Here $ExpIntegralEi[z]$ gives the exponential integral function $Ei[z]$, where $\text{Ei}(z)=-\int_{-z}^{\infty } \left.e^{-t}\right/t \, dt$.

This solution can be illustrated by the graph (we assume that $\alpha=1,\, \beta=1\, \kappa=1, \, \rho_m=0$):

\begin{figure}[!h]
\begin{minipage}[h]{0.49\linewidth}
\center{\includegraphics[angle=0, width=1\textwidth]{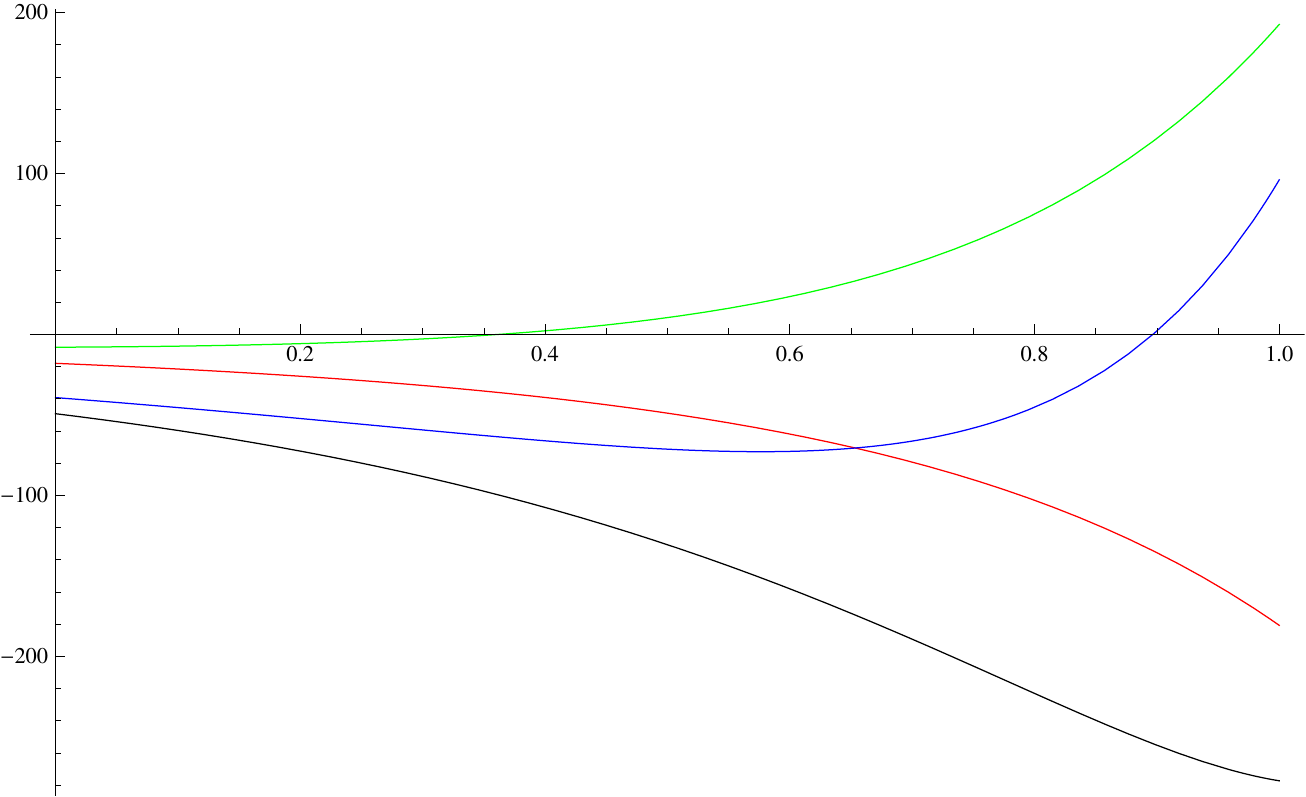}\\a}
\end{minipage}
\hfill
\begin{minipage}[h]{0.49\linewidth}
\center{\includegraphics[angle=0, width=1\textwidth]{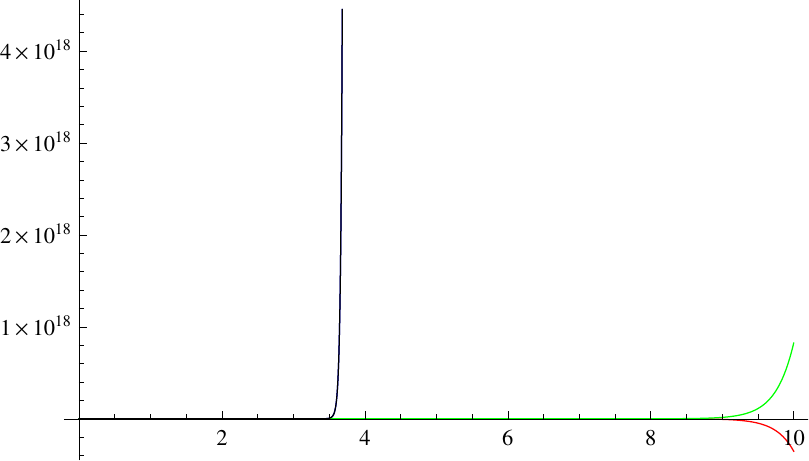}\\b}
\end{minipage}
\caption{Plot of $f(t)$, a) (t,0,1), b) (t,0,10) (Green line - $C_2=0,\,C_1=10$, red line - $C_2=0,\,C_1=0$, blue line - $C_2=10,\,C_1=10$, black line - $C_2=10,\,C_1=0$).}
\end{figure}

Figures show that at the initial time all the solutions have the same behavior (dominated by the first term in (\ref{sol1})), then
begins to play the role of the second term ($C_1\ne 0$), and finally turned the third term ($C_2\ne 0$), this is clearly seen in the figure 1.

\begin{figure}[-!h]
\begin{center}
\begin{minipage}[h]{0.4\linewidth}
\includegraphics[angle=0, width=1\textwidth]{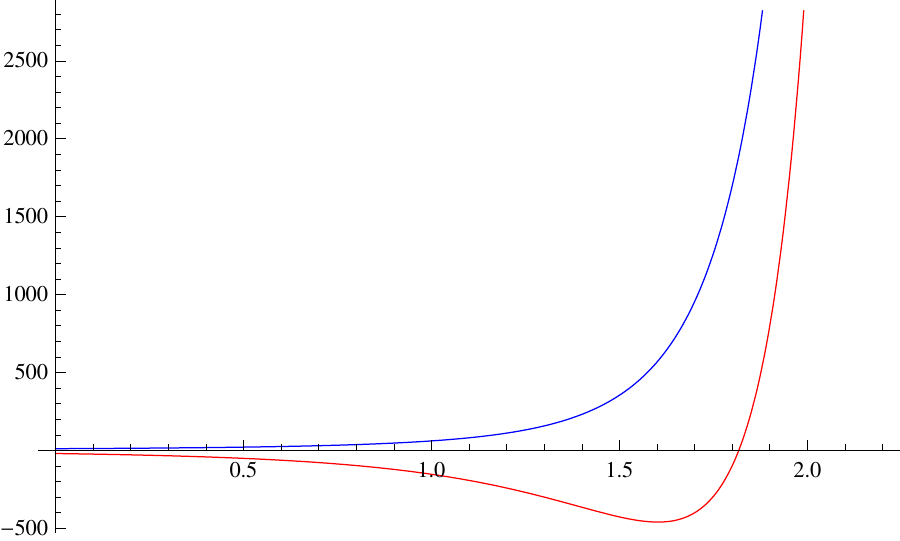}
\caption{Plot of $f(t)$, (t,0,2.2) (Red line - $C_2=1,\,C_1=1$, blue line - $f(t)=4e^{e^t}$).}
\end{minipage}
\hfill
\begin{minipage}[h]{0.4\linewidth}
\includegraphics[angle=0, width=1\textwidth]{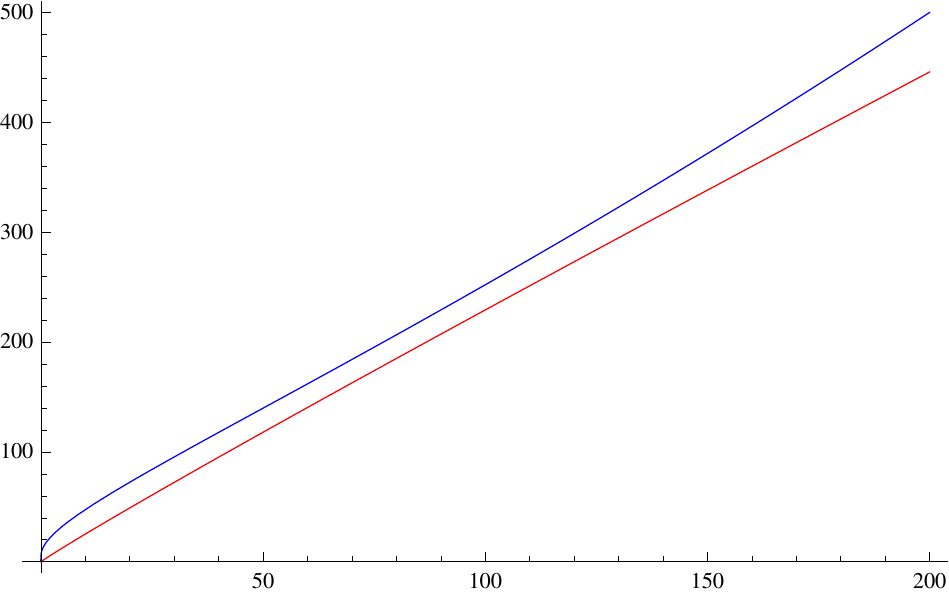}
\caption{Plot of $f(G)$, (G,0,200) (Red line - $C_2=-10,\,C_1=100$, blue line - $f(G)=8G^{1/8}e^{G^{1/4}}$).}
\end{minipage}
\end{center}
\end{figure}

We can consider the dependence of the $F(G)$. Obviously not possible to do this, but for large $H$,
we can assume that $G=24 H^4$. From (\ref{sol1}) one can see that there are 3 terms: the first dominates at small $G$, then by $30<G<100$ included the last term
($C_2\ne 0$ and $f(G)\sim G^{1/8}e^{G^{1/4}}$, figure 3),  then included the second term ($C_1\ne 0$ and $f(G)\sim G$, figure 4a) and then
$G->\infty$ (figure 4b) again dominated by the last term.

\begin{figure}[!h]
\begin{minipage}[h]{0.49\linewidth}
\center{\includegraphics[angle=0, width=1\textwidth]{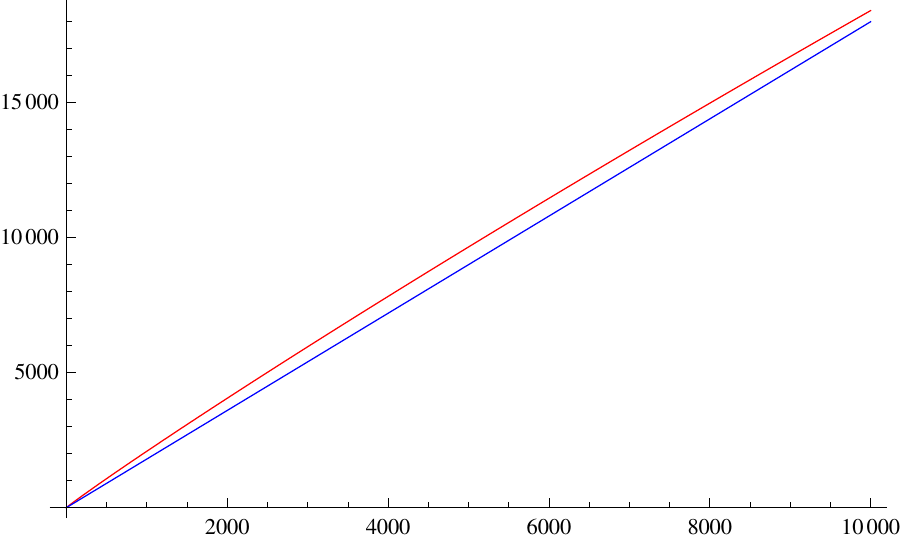}\\a}
\end{minipage}
\hfill
\begin{minipage}[h]{0.49\linewidth}
\center{\includegraphics[angle=0, width=1\textwidth]{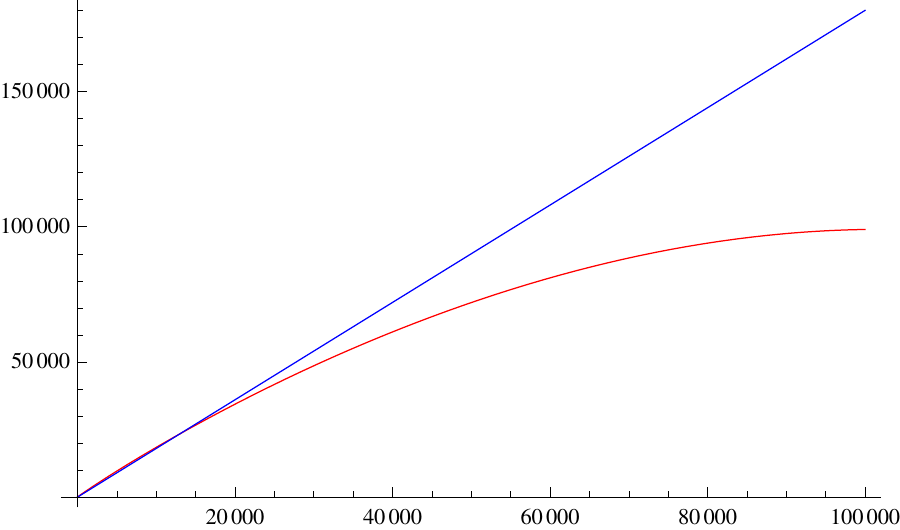}\\b}
\end{minipage}
\caption{Plot of $f(G)$ a) (G,0,10000), b) (G,0,100000) (Red line - $C_2=-10,\,C_1=100$, blue line - $f(G)=1.8G$).}
\end{figure}

This behavior is due to the fact that the last term consists of two parts with different signs. For small values of $G$ terms rather different, and with increasing $G$, they become the same order and the difference becomes smaller than other terms. When $G$ tends to infinity, the difference between them begins to grow faster than the other terms.

By selecting different values of the constants $C_1$ and $C_2$ can be different depending $f$ from $G$.
For example, we construct the graph for the following $f(G)$ values of the constants of integration: $C_2=1$, $-20<C_1<20$ (figure 5).

\begin{figure}[!h]
\begin{center}
\includegraphics[angle=0, width=0.5\textwidth]{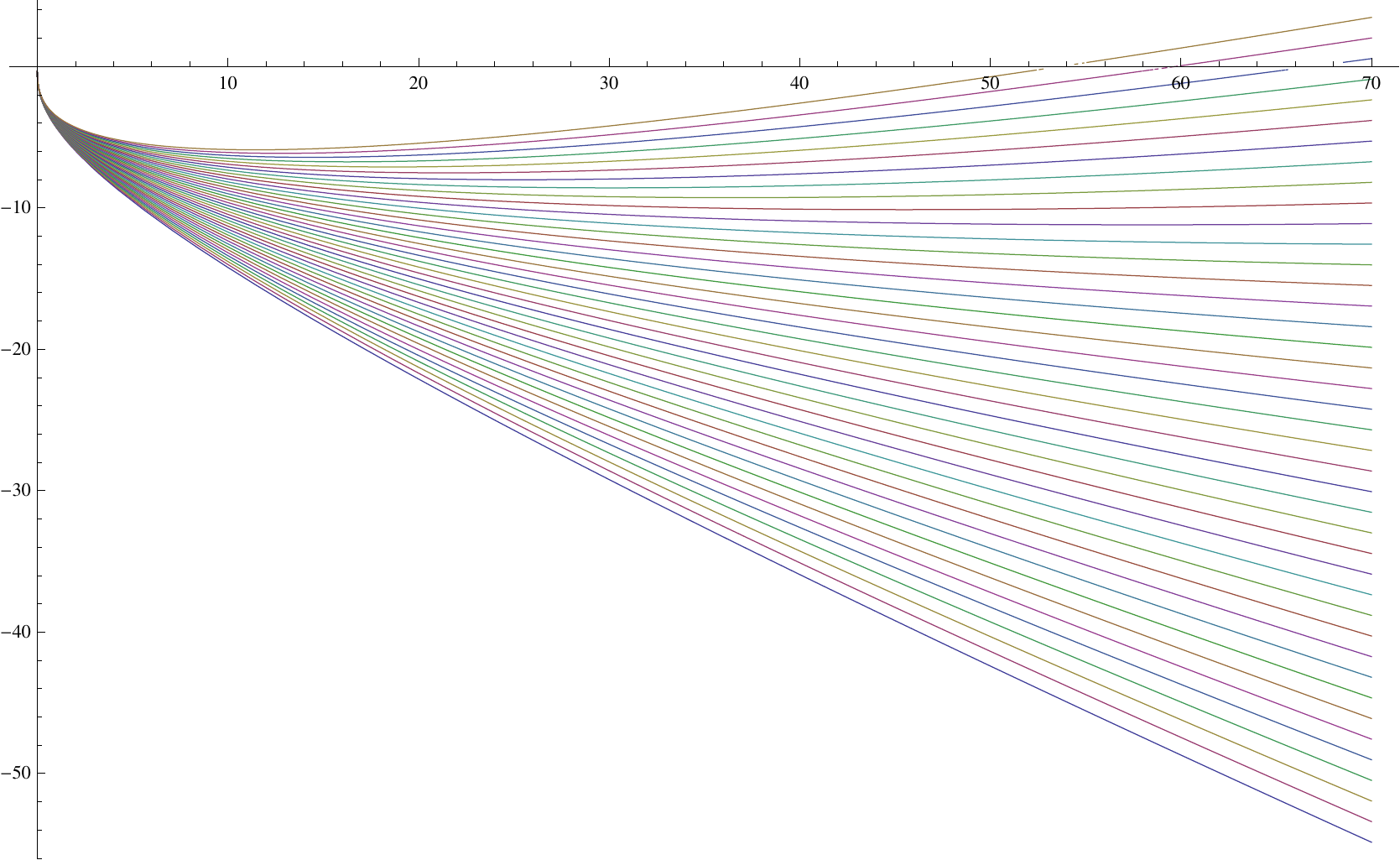}
\end{center}
\caption{Plot of $f(G)$, (G,0,70) ($C_2=1$, $-20<C_1<20$).}
\end{figure}

Consider now the $F(G,R)$ - gravity. Let us solve the equations (\ref{3.1})
\begin{eqnarray}
&&f_2(H)=\frac{1}{\left(4 \alpha (2+\alpha (2+(-5+\alpha) \alpha)) \beta^4\right)}\\
&&\left(\beta e^{\frac{H}{\beta}} H \left(\beta^2+4 \beta H-H^2\right) C_2+H \left(2 \beta^3+2 \beta^2 H-5 \beta H^2+H^3\right)
\left(4 C_1+C_2 \text{ExpIntegralEi}\left[\frac{H}{\beta}\right]\right)\right)\nonumber
\end{eqnarray}

\begin{eqnarray}
&&f_1(H)=\frac{1}{2 \alpha^3 (1+\alpha) \beta^4}\\
&&\left(-\alpha^2 \beta e^{\frac{H}{\beta}} H \left(3 \beta^2+2 \beta H+H^2\right) C[2]+H^3 (\beta+H) \left(2 C[1]+\alpha^2 C[2] \text{ExpIntegralEi}\left[\frac{H}{\beta}\right]\right)\right)\nonumber
\end{eqnarray}

Using the
\begin{equation}
H= \frac{1}{12} \left(-3 \beta+\sqrt{9 \beta^2+12 R}\right).
\end{equation}
We obtain the dependence of the function $f_2$ from $R$

\begin{eqnarray}
&&f_2(R)=
\left(-3 \beta+\sqrt{9 \beta^2+12 R}\right) \left(-6 \beta e^{\frac{-3 \beta+\sqrt{9 \beta^2+12 R}}{12 \beta}} \left(2 R+3 \beta \left(\beta-3 \sqrt{9 \beta^2+12 R}\right)\right) C_2+\right.\nonumber\\
&&\left(117 \beta^3-69 \beta R+57 \beta^2 \sqrt{9 \beta^2+12 R}+R \sqrt{9 \beta^2+12 R}\right)\\
 &&\left.\left(4 C_1+C_2 \text{ExpIntegralEi}\left[\frac{-3 \beta+\sqrt{9 \beta^2+12 R}}{12 \beta}\right]\right)\right)\nonumber
\end{eqnarray}

We can consider the dependence of the $f_1(G)$. Obviously not possible to do this, but for large $H$,
we can assume that $G=24 H^4$. Then

\begin{eqnarray}
&&f_1(G)=-6 \text{al}^2 \beta e^{\frac{G^{1/4}}{2^{3/4} 3^{1/4} \beta}} \left(12 \sqrt{3} \beta^2+4\ 6^{1/4} \beta G^{1/4}+\sqrt{2} \sqrt{G}\right) G^{1/4} C_2+\nonumber\\
&&+\left(6 \sqrt{2} \beta G^{3/4}+6^{3/4} G\right) \left(2 C_1+\text{al}^2 C_2 \text{ExpIntegralEi}\left[\frac{G^{1/4}}{2^{3/4} 3^{1/4} \beta}\right]\right)
\end{eqnarray}

\section{Power law solutions}

Consider a power-law solutions. Choose the Hubble parameter as
\begin{equation}
H(t)=\frac{\alpha}{t_s-t}.
\label{big_rip}
\end{equation}
Then the scale factor is given by
\begin{equation}
a(t)= \frac{C}{(t_s-t)^\alpha}.
\label{big_rip1}
\end{equation}

This scale factor can meet the various phases of expansion, which are characterized by different values of the equation of state (EoS) parameter
$w_{DE}$.
To easily find the value $w_{DE}$ for (\ref{big_rip1}). The equations FRW have the form:
\begin{equation}
\rho=\frac{3}{\kappa^2} H^2, \,\,\, p=-\frac{1}{\kappa^2}\left( 2 \dot{H}+3H^2\right).
\end{equation}
Comparing with the (\ref{GB13}), we can write formally the same equations by removing all unnecessary to the effective pressure and energy density;:
\begin{eqnarray}
\rho_{eff}&=&G
f_{G}-f(G)-24\dot{G}H^{3}f_{GG}+\rho_{m}\nonumber\\
p_{eff}&=&8H^2\ddot{f}_{G}+16H(\dot{H}+H^2)\dot{f}_G+f-G f_G+p_m\,.
\label{eff}
\end{eqnarray}
So
\begin{equation}w_{DE}=\frac{p_{eff}}{\rho_{eff}}=-\frac{1}{3}\frac{2\dot{H}+3H^2}{H^2}.
\end{equation}
For (\ref{big_rip}) have
$w_{DE}=-1/3 (2/\alpha+3)$.

Consider the different values of $w_{DE}$. If $w_{DE}$ = -1 this could be consistent if the
dark energy is a cosmological constant and therefore our universe seems to approach to asymptotically
de Sitter universe. This value is not possible for the chosen scale factor.
This value is the min in the permissible region, but the possibility that the current acceleration
could be quintessence type, in which $w_{DE}$ > -1, or phantom type, in which $w_{DE}$ < -1, is not completely excluded.
For quintessence type, the value of $\alpha$ is negative. For phantom type, the value of $\alpha$ is positive.

It is easy to write of the equation (\ref{FFF1}) for this cases as an equation for $f(t)$
\begin{eqnarray}
0&=&-4 (1+\alpha) \left(3 \alpha^2-\kappa^2 \rho_0 (t_s-t)^2\right)+\nonumber\\
&+&\kappa^2 (t_s-t)^2 \left(-4 (1+\alpha) f(t)-(t_s-t) \left(-(6+\alpha) f'(t)+(t_s-t) f''(t)\right)\right)
\end{eqnarray}
where $\rho_0=const$.

\begin{equation}
f(t)=
 \rho_0-\frac{6 \alpha^2 (1+\alpha)}{(-1+\alpha) \kappa^2 (t_s-t)^2}+(t_s-t)^{-1-\alpha} C_1+ \frac{C_2}{(t_s-t)^4}
\end{equation}

We now express $t$ through $G$
\begin{equation}
t_s-t=\frac{2^{3/4} 3^{1/4} \left(\alpha^3 +\alpha^4 \right)^{1/4}}{G^{1/4}}.
\end{equation}
And we obtain the following form of the function $f(G)$

\begin{equation}
f(G)=-\frac{\sqrt{\frac{3}{2}} \sqrt{\alpha^3 (1+\alpha) G}}{(-1+\alpha) \alpha \kappa^2}+\rho_0-i^{-2 \alpha} 2^{-\frac{3}{4} (1+\alpha)} 3^{\frac{1}{4} (-1-\alpha)} G^{\frac{1+\alpha}{2}} \left(\alpha^3 (1+\alpha) G\right)^{\frac{1}{4} (-1-\alpha)} C_1
\end{equation}

It is seen that if $\alpha=3$ the equation is considerably simplified
\begin{equation}
-\frac{3 \sqrt{G}}{\sqrt{2} \kappa^2}+\rho_0.
\end{equation}

Now consider the space filled with dust
$\rho_m=\frac{\rho_0}{a^3}$.
\begin{eqnarray}
&&f(G)=-\frac{\sqrt{\frac{3}{2}} \sqrt{\alpha^3 (1+\alpha) G}}{(-1+\alpha) \alpha \kappa^2}+\frac{2^{2+\frac{9 \alpha}{4}} 3^{3 \alpha/4} (1+\alpha) G^{-3 \alpha/2} \left(\alpha^3 (1+\alpha) G\right)^{3 \alpha/4} p_0}{\left(4+19 \alpha+12 \alpha^2\right) \text{C}^3}-\nonumber\\
&&-2^{-\frac{3}{4} (1+\alpha)} 3^{\frac{1}{4} (-1-\alpha)} e^{-i \alpha \pi } G^{\frac{1+\alpha}{2}} \left(\alpha^3 (1+\alpha) G\right)^{\frac{1}{4} (-1-\alpha)} C_1
\end{eqnarray}

It is seen that if $\alpha=3$ the equation is considerably simplified
\begin{equation}
f(G)=-\frac{3 \sqrt{G}}{\sqrt{2} \kappa^2}+\frac{644972544\ 2^{1/4} \rho_0}{169 C^3 G^{9/4}}.
\end{equation}

Consider now the $F(R,G)$ - gravity (\ref{3.1}). Equations take the form
\begin{equation}
0={4 (1+\alpha) f_1-(t_s-t) \left((6+\alpha) f_1'-(t_s-t) f_1''\right)},
\end{equation}
\begin{equation}
0={(2+4 \alpha) f_2-(t_s-t) \left((4+\alpha) f_2'-(t_s-t) f_2''\right)}.
\end{equation}

The solution of these equations have the form
\begin{equation}f_1= \frac{(t-t_s)^{3-\alpha} C_1+C_2}{(t_s-t)^4},\,\,t_s-t=-\frac{2^{3/4} 3^{1/4} \left(\alpha^3 G+\alpha^4 G\right)^{1/4}}{\sqrt{G}}
\end{equation}
\begin{equation}f_2=(t-t_s)^{\frac{1}{2} \left(-3-\alpha-\sqrt{1+(-10+\alpha) \alpha}\right)} \left(C_1+(t-t_s)^{\sqrt{1+(-10+\alpha) \alpha}} C_2\right),\,\,
t_s-t=\frac{\sqrt{6} \sqrt{\alpha(1+2 \alpha)}}{\sqrt{R}}.
\end{equation}

Or finally get
\begin{equation}
f_1(G)=C_1 G^{(1+\alpha)/4} ,
\end{equation}
\begin{equation}
f_2(R)=R^{\frac{1}{4} \left(3+\alpha+\sqrt{1+(-10+\alpha) \alpha}\right)} \left(C_1+ R^{-\frac{1}{2}\sqrt{1+(-10+\alpha) \alpha}} C_2\right).
\end{equation}

\section{de Sitter solutions}

De Sitter solutions are described by an exponential expansion of the
Universe, where the Hubble parameter and the scale factor are given
by
\begin{equation}
H(t)=H_0\rightarrow a(t)=e^{H_0 t}\ ,
\label{D8}
\end{equation}
where $H_0$ is a constant. This kind of solutions are very
important, as the observations suggest that the expansion of our
Universe behaves approximately as de Sitter. It has been shown in
Ref. \cite{Tret} that de Sitter points are critical points
in $f(R)$ gravity. It is straightforward to see that this is also
the case in $R+f(G)$ gravity.
\begin{equation}
0=-\frac{3}{\kappa^2}H_0^2+G_0f_G(G_0)-f(G_0)\ .
\label{D9}
\end{equation}
Here, $G_0=24H_0^2$ and we have ignored the contribution of matter.
Then, we have reduced the differential equation to an
algebraic equation that can be resolved by specifying a function
$f(G)$.

 Consider the equation (\ref{D9}) as a differential equation for $G$:
\begin{equation}
0=-\sqrt{\frac{3}{2}}\frac{\sqrt{G}}{2\kappa^2}+Gf_G(G)-f(G)\ .
\label{D10}
\end{equation}
In this case, the function $F(G)$ takes the form
\begin{equation}
f(G)=-\frac{\sqrt{\frac{3}{2}} \sqrt{G}}{\kappa^2}.
\end{equation}

We see that the function of the form $f(G)=c \sqrt{G}$ describes the power solutions and de Sitter solution ($c=-\sqrt{\frac{3}{2}}\frac{1}{\kappa^2}$).
The general solution has the form
\begin{equation}
a(t)=(t-\alpha t-C_1)^{\frac{1}{1-\alpha}} C_2,
\end{equation}
here $C_1$ and $C_2$ -constants of integration. The case when $\alpha = 1$ corresponds to the de Sitter solution.

For $F(G, R)$ gravity similarly, we have the following equations
\begin{equation}
0=-\frac{1}{2}\left( G_0 F_G (G_0)-F(G_0, R_0)\right)+3H_0^2 F_R(R_0)\ .
\end{equation}
If we consider the case $F(G,R)=f_1(G)+f_2(R)$ is not difficult to find a possible solution $F(R,G)=R^2$.

\section{Little Rip in Gauss-Bonnet gravity with dilaton}

We can find a Little Rip solution for Gauss-Bonnet gravity with dilaton  for spaces of dimension $D+4$.
Consider space of $D=p+q+1$ dimensions with two maximally symmetric
subspaces: $p$-dimensional and $q$-dimensional.
We write the Lagrangian as
\begin{equation}\label{EGBd-lagr}L=\frac{R}{2\kappa^2} -\frac{1}{2} \partial_\mu\varphi\partial^\nu\varphi-V(\varphi)+
\varepsilon(\varphi)G.\end{equation}

Here $\varepsilon(\varphi)$, $V(\varphi)$ are functions of dilaton
$\varphi$. We choose them as
\begin{equation}\label{epsilonV}\varepsilon(\varphi)=\beta
e^{-\gamma\varphi},\quad V(\varphi)=\alpha e^{\gamma\varphi}.\end{equation}
Variating the action with Lagrangian (\ref{EGBd-lagr}) we get field
equations \cite{mak1}.
One of the possible solutions of these equations will be
\begin{equation}\label{scale-1}a(t)=a_0
\exp\left\{\dac{2u_1 c_0}{u_3}e^{u_3 t/2}\right\},\end{equation}
\begin{equation}\label{scale-2}b(t)=b_0
\exp\left\{\frac{2u_2 c_0}{u_3}e^{u_3 t/2}\right\}\end{equation}
where $a_0, b_0$ and $c_0$ are arbitrary positive constants.
$u_1, u_2, u_3$ are constants satisfying the following equations

\begin{equation}\label{eq0-exp-exp}\begin{array}{l} 3{u_1}^2+\dac{1}{2}q_1 {u_2}^2+3q u_1u_2+
\dac{1}{2}{ u_3}^2-\dac{1}{2}\alpha+\dac{1}{2}\beta\{36q_1{u_1}^2{u_2}^2+
q_3{u_2}^4+\\
\quad{}+12u_1u_2(2q{u_1}^2+q_2{u_2}^2)\}-\vphantom{\dac{\dac{1}{2}}{2}}
2\beta\gamma u_3\{18q{u_1}^2 u_2+9q_1 u_1{u_2}^2+
6q{u_1}^3+q_2{u_2}^3\}=0;\vphantom{\dac{\dac{1}{2}}{2}}\end{array}\nonumber
\end{equation}

\begin{equation}\label{eqij-exp-exp}\begin{array}{l}-3{u_1}^2-\dac{1}{2}(q+1)q{u_2}^2-
\gamma u_3\(u_1+\dac{q}{2}u_2\)-2qu_1u_2+\dac{1}{2}{ u_3}^2+
\dac{\alpha}{2}-\dac{1}{2}\beta\{-4q(3q+1)\gamma u_3u_1{u_2}^2-\\
\quad{}-
28q\gamma u_3{u_1}^2u_2-\vphantom{\dac{\dac{1}{2}}{2}}
2(q+2)q_1\gamma u_3{u_2}^3-4\gamma^2{ u_3}^2{u_1}^2-2q_1\gamma^2{ u_3}^2{u_2}^2+
4q(5q-3){u_1}^2{u_2}^2+\vphantom{\dac{\dac{1}{2}}{2}}\\
\quad{}+
8qq_1u_1{u_2}^3+16q{u_1}^3u_2+(q+1)q_2{u_2}^4-16\gamma u_3{u_1}^3-
8q\gamma^2{ u_3}^2u_1u_2\}=0;\vphantom{\dac{\dac{1}{2}}{2}}\end{array}\nonumber\end{equation}

\begin{equation}\label{eqab-exp-exp}\begin{array}{l}-\dac{1}{2}q_1{u_2}^2-6{u_1}^2-\dac{q-1}{2}\gamma u_3u_2-
\dac{3}{2}\gamma u_3u_1-3(q-1)u_1u_2+\dac{1}{2}{ u_3}^2+\dac{\alpha}{2}-{}\\
\quad{}-\vphantom{\dac{\dac{1}{2}}{2}}\dac{1}{2}\beta
\{-2(q+1)(q-1)_2\gamma u_3{u_2}^3-
6(q-1)(3q-2)\gamma u_3u_1{u_2}^2-
60(q-1)\gamma u_3{u_1}^2u_2-\\
\quad{}-\vphantom{\dac{\dac{1}{2}}{2}}60\gamma u_3{u_1}^3-2(q-1)_2\gamma^2{ u_3}^2{u_2}^2-12\gamma^2{ u_3}^2{u_1}^2+
q_3{u_2}^4+\vphantom{\dac{\dac{1}{2}}{2}}24(q-1)(2q-3){u_1}^2{u_2}^2+\\
\quad{}+12(q-1)(q-1)_2u_1{u_2}^3+72(q-1){u_1}^3u_2+
\vphantom{\dac{\dac{1}{2}}{2}}24{u_1}^4-12(q-1)\gamma^2{ u_3}^2u_1u_2\}=0;\end{array}\nonumber\end{equation}

\begin{equation}\label{eqphi-exp-exp}\begin{array}{l}-\gamma { u_3}^2-2(3u_1+q u_2) u_3+
\beta\gamma
\{12q(4q-2){u_1}^2{u_2}^2+(q+1)q_2{u_2}^4+72q{u_1}^3u_2+
12qq_1u_1{u_2}^3+\\
\quad{}+\vphantom{\dac{\dac{1}{2}}{2}}18q_1\gamma u_3u_1{u_2}^2+36q\gamma u_3{u_1}^2u_2+
12\gamma u_3{u_1}^3+24{u_1}^4+2q_2\gamma u_3{u_2}^3
\}+\alpha\gamma=0.\vphantom{\dac{\dac{1}{2}}{2}}\end{array}\nonumber\end{equation}

Therefore it is necessary to find solutions of
this equations satisfied conditions
$$u_1>0,\quad u_2<0,\quad  u_3>0.$$
This is easily done, for example
$$\begin{array}{lllll}q=1,\quad& \alpha=1,\beta=1,\gamma=1,\quad&  u_3=0.383,\quad&
u_1=0.378,\quad& u_2=-1.32.\vphantom{dac{1}{2}}\\
q=2,\quad& \alpha=0.01,\beta=100,\gamma=0.01,\quad&
 u_3=5.09\cdot10^{-5},\quad& u_1=0.0175,\quad& u_2=-0.0803.\vphantom{dac{1}{2}}\\
q=3,\quad& \alpha=0.001,\beta=100,\gamma=0.001,\quad&
 u_3=2.33\cdot10^{-4},\quad& u_1=0.0306,\quad&
u_2=-0.0788.\vphantom{dac{1}{2}}\end{array}$$

\section{Conclusions}

 In this paper we construct a series of cosmological models which describe the accelerated expansion of the universe.
 The models in framework of the modified Gauss-Bonnet gravity are constructed by the reconstruction method.
 Constructed models are of the effective phantom type and do not lead to a singularity in the future (Little rip models).

 An explicit form of the action leading to such a model within the framework of gravity $R+f(G)$ and $f_2(R)+f_1(G)$ is obtained.
 The solution contains special function, which can be removed by choosing the constant of integration.
The resulting action consists of three terms, each of which plays a role at the definite stage of the  evolution of the Universe.
Since the last term represents the difference between the functions of the same order, and this order of these functions is maximal, it has double the lead - with the average values and tends to infinity. In addition we construct a model describing the Little Rip in the multivariate Gauss-Bonnet theory with dilaton, when visible, and additional spaces are maximally symmetric.

Little Rip models provide an evolution for the universe intermediate between asymptotic de Sitter expansion and models with a big rip singularity. Therefore, the method of reconstruction models were built describing the de Sitter space and space with the metric \ref{big_rip1}, which refers to the phantom type and has a singularity of the future.

For all constructed models $w_{DE}<-1$ (except deStter model) and therefore they belong to the phantom type.
For models of the effective phantom energy conditions are always broken. Such behavior is typical for the Little Rip models \cite{LR1}

For Little Rip models with Hubble parameter (\ref{h1}), we obtain
$$ \rho+p=-\frac{2 \alpha \beta^2 e^{\beta t}}{\kappa^2}.$$

 We can continue to explore these theories and consider stability of solutions. However, as shown by work \cite{LR} Little Rip solutions are stable and can be expected and received from our paper solutions will have the same property.

\end{document}